\documentclass[12pt,oneside]{article}
\def\doublespace {\smallskipamount=7.5pt plus2pt minus2pt
                  \medskipamount=15pt plus4pt minus4pt
                  \bigskipamount=30pt plus8pt minus8pt
                  \normalbaselineskip=30pt plus0pt minus0pt
                  \normallineskip=2pt
                  \normallineskiplimit=0pt
                  \jot=7.5pt
                  {\def\smallskip {\vskip\smallskipamount}}
                  {\def\medskip   {\vskip\medskipamount}}
                  {\def\bigskip   {\vskip\bigskipamount}}
                  {\setbox\strutbox=\hbox{\vrule 
                    height21.0pt depth9.0pt width 0pt}}
                  \parskip 15.0pt
                  \normalbaselines}

\def\sect #1{\setcounter{equation}{0}}

\begin{document}

\title{Gravitational Collapse and Cosmological Constant}

\author{S. S. Deshingkar${}^{1}$\thanks{shrir@relativity.tifr.res.in}, 
S. Jhingan${}^{2}$\thanks{wtpsaxxj@lg.ehu.es}, 
A. Chamorro${}^{3}$\thanks{wtpchbea@lg.ehu.es}, and
P. S. Joshi${}^{4}$\thanks{psj@tifr.res.in} \\
${}^{2}$Tata Institute of Fundamental Research, \\
Homi Bhabha Road, Mumbai 400005, INDIA \\
${}^{2}$Departamento de Fisica Teorica, 
Universidad del Pais Vasco,\\ 
Apdo. 644, E-48080 Bilbao, Spain. \\}

\maketitle
\newpage

\begin{abstract}
We consider here the effects of a non-vanishing cosmological term 
on the final fate of a spherical inhomogeneous collapsing dust cloud. 
It is shown that depending on the nature of the initial data from which 
the collapse evolves, and for a positive value of the cosmological
constant, we can have a globally regular evolution where a bounce develops
within the cloud. We characterize precisely the initial data causing 
such a bounce in terms of the initial density and velocity profiles for
the collapsing cloud. In the cases otherwise, the result of collapse is
either formation of a black hole or a naked singularity resulting 
as the end state of collapse. We also show here that a positive 
cosmological term can cover a part of the singularity spectrum which is 
visible in the corresponding dust collapse models for the same 
initial data.
\end{abstract}
\doublespace
\section{Introduction} 

The gravitational collapse of a dust cloud has been studied extensively 
in the literature~\cite{Jhingan}, especially in the light of the cosmic
censorship conjecture of Penrose~\cite{Penrose}. There are no globally 
regular solutions 
in these models, and a singularity always develops in an initially collapsing
configuration. These singularities could be naked, or hidden behind an 
event horizon, depending on the nature of the initial data
functions representing the density and velocity profiles for the 
collapsing cloud~\cite{Jhingan}, from which the collapse evolves. 
It is also known that the shell-focusing
singularities in dust models are strong curvature
singularities~\cite{Shrirang}. In this sense these are physically
genuine singularities and the possibility of extension of the spacetime 
through the same does not arise.

In investigations such as above, one generally uses the Einstein equations
with a vanishing cosmological term. Recent observations however give 
us an evidence that a large part of the universe is possibly dominated 
by an energy component with negative pressure~\cite{Perl, Dekel}. 
Among various possibilities available~\cite{Starobinsky}, one candidate 
that is often considered is a cosmological constant~\cite{Sahni}, $\Lambda$, 
or vacuum energy density corresponding to a positive sign of $\Lambda$. 
Such a cosmological constant would represent a spatially uniform 
energy density distribution, which is time-independent, and its positive 
value acts as a globally repulsive force field. Also, the cold dark matter
models with a substantial component supplied by the cosmological term
are among the models which
best fit the observational data~\cite{Ost}. Therefore it is of
value to revive the cosmological term, as a constant or even as a time
varying quantity, in the Einstein equations. With such a perspective 
in mind we have investigated here the gravitational collapse, 
and structure of the singularity in the spherically symmetric dust 
models with the presence of a non-vanishing cosmological constant.

Dust models with a cosmological term are also known in the
literature~\cite{Krasinski}. These models can be matched with the
Schwarzschild-de Sitter spacetime at the boundary of the 
cloud~\cite{Kotler}. But the
studies so far have been restricted to special cases, whereas we 
analyze here the Lemaitre-Tolman-Bondi (LTB) collapse models \cite{LTB}
with a cosmological constant, which provide the general solution
to Einstein equations with dust as the source term. 
While studying these models, the weak energy condition, 
i.e. $T_{ij}V^iV^j \geq 0$, for all
non-spacelike vectors $V^i$, is assumed everywhere in spacetime
for the matter, which would be a physically reasonable condition to 
assume for the case of a collapsing cloud.

In the next Section 2 we introduce the basic model and analyze the
Einstein equations to check whether we can have globally regular solutions, 
as shown to exist in the case of homogeneous density
dust solutions with a positive cosmological constant~\cite{Markovic}. 
We also derive the general solution to the Einstein equations in 
the case corresponding to the  marginally bound case in the LTB models 
(i.e. the case when energy function $f=0$),
and discuss the condition for avoiding shell-crossings. 
In Section 3 we investigate the structure of the singularity by studying
(O)outgoing (Ra)dial (N)ull (Ge)odesics (ORANGEs) near the same. The
bearing of the initial conditions - in the presence of the cosmological
term - on whether the final state results in a black hole or a naked
singularity is then analyzed. Section 4 provides the conclusions.


\section{Dust collapse with $\Lambda$ term}

In this section we first study the basic set of Einstein 
equations and the regularity conditions for collapse. Then we will 
look for the possibility of regular solutions when there is a 
non-vanishing cosmological term present,  wherein an initially
collapsing cloud rebounces at some later epoch 
so that singularity does not form.
This corresponds to the occurrence of three phases, which are,
collapse, reversal and subsequent dispersal. 
Finally, we give here a general solution with non-zero $\Lambda$
term for the marginally bound case.

\subsection{Basic equations and regularity}

The model for a self-gravitating, spherically symmetric,
inhomogeneous dust cloud with cosmological constant is given 
by the metric,
\begin{equation} ds^2=-dt^2 +
\frac{{R'}^2}{1+f(r)}dr^2 +R^2(d\theta^2+{\sin}^2 \theta d\phi^2),
\label{metric} 
\end{equation} 
where $(t,r,\theta,\phi)$ is a comoving coordinate system. 
Here $f(r) > -1 $, and $r$ and $R$ are the
shell-labeling comoving coordinate, and the physical area radius  
respectively. In our notation the dot and the prime denote partial 
derivatives with respect to $t$ and $r$ respectively.
The equation of
state for matter in the interior of the cloud is that of dust 
(corresponding to the approximation of a fluid with negligible pressures),
and the stress-energy tensor is given by, 
$$
T^i_j = \epsilon (r,t) \delta^i_t \delta^t_j, 
$$
where $\epsilon (r,t)$ is the energy density of matter. We assume the 
matter to satisfy the weak energy
condition, which implies that $\epsilon (r,t) \geq 0$; it is 
equivalent to the strong energy condition as the principal pressures 
are zero. The Einstein equations in the presence of a cosmological 
constant can be written as,
\begin{equation}
\dot R^2= \frac{F(r)}{R} + f(r) + \frac{\Lambda}{3} R^2,
\label{dyna}
\end{equation}
\begin{equation}
\epsilon(t,r)= \frac{F'}{ R^2R'}.
\label{sing}
\end{equation}
We will be mainly considering the collapse situation (i.e. $\dot R <0$).
The dust cloud is characterized by two free
functions in general, representing the total mass ($F(r)$), and the
energy ($f(r)$), inside the shell labeled by comoving coordinate $r$. 
The cosmological term
$\Lambda$ can in principle be of either sign, however the recent       
observations as indicated above seem to favour the positive sign.
The above equation for $\dot R$ can be in principle
integrated, and after integration one gets a constant of integration.
One can fix the constant of integration by using the scaling freedom.
We fix this by setting $R=r$ on the initial hypersurface ($t=0$). 
The two free functions $F(r)$ and $f(r)$
can be fixed by prescribing the initial density and velocity profiles 
through (\ref{sing}) and (\ref{dyna}) respectively. 
We assume these free functions to have the form
\begin{equation}
\begin{array}{c}
F(r)= F_0r^3+F_nr^{3+n} +higher \; order \; terms,\\
f(r)= f_0r^2+f_{n_1}r^{2+n_1} +higher \; order \; terms,
\end{array}
\label{initialdata}
\end{equation}
where the choice of the first non-vanishing term is
made in order to have a regular initial data on the initial
surface $t=0$ from which the collapse evolves~\cite{Jhingan}. 
From~(\ref{sing}) it is clear that we can have both shell-crossing
($R'(t_{sc}(r),r)=0$), and shell-focusing singularities
($R(t_{sf}(r),r)=0$) in these spacetimes, depending on the dynamics of
the shells given by~(\ref{dyna}). In this paper we will only consider cases 
where there are no shell-crossings, because these are generally not
considered to be genuine singularities, and as our main
interest is in studying the nature of physical singularity corresponding 
to $R=0$ where the matter shells shrink to zero radius. 
This puts some restrictions on our 
initial data which we will discuss later.

The metric exterior to the collapsing cloud has the Schwarzschild-de Sitter 
form,
\begin{equation}
ds^2 = -g dt^2 + g^{-1} dr^2 + r^2 (d\theta^2+{\sin}^2 \theta d\phi^2) ,
\label{exterior}
\end{equation}
where $g = 1-2M/r-\Lambda r^2/3$. Matching the solutions~(\ref{metric})
and~(\ref{exterior}) at the boundary $r_b$ of the collapsing dust cloud
one obtains $2M = F(r_b)$~\cite{Santos, Markovic}. It is known that the
boundary can be made to bounce from an initially collapsing phase by
choosing appropriate initial mass and cosmological term, as 
we see bellow.
This behaviour can also be understood by analyzing the curve of allowed
motion~(\ref{dyna}) (for details see,~\cite{Santos},~\cite{Markovic}).

\subsection{Regular solutions and rebounce}

First we look for the possibility of having globally regular 
solutions due to the presence of a cosmological constant.
The equation~(\ref{dyna}), governing the dynamics of
collapsing shells, can be conveniently written as
\begin{equation}
\dot R^2= \frac{V(R,r)}{3R} .
\label{Pot}
\end{equation}
Here $V(R,r)$, defined as
\begin{equation}
V(R,r) = 3 F(r) + 3 f(r) R + \Lambda R^3 ,
\label{pot}
\end{equation}
is an analogue of the Newtonian effective potential governing motion
of the shells. The allowed region of motion corresponds to $V \geq
0$ as $\dot R^2$ has to be greater than zero.

If we start from an initially collapsing state, we will have a 
rebounce if 
we get $\dot R=0$ for a given shell before that shell becomes singular.
This can happen only if the equation
$$
3R\dot R^2 = V(R,r)=0
$$
has two real positive roots. 
In what follows it is convenient to define a quantity
$\zeta(t,r) = R/r$. The equation (\ref{pot}) can be rewritten as
\begin{equation}
V = 3 F  + 3 f \zeta r + \Lambda \zeta^3 r^3 .
\label{Vzeta}
\end{equation}
which is a cubic equation with three roots in general. 
>From the theory of cubic equations, if all the three roots of the equation 
above are real then at least one of them has to be positive and at least 
one negative. Note that $V(R=0,r>0)=3F>0$. Hence, any regular region between
$R=0$ and the first zero of eqn.~(\ref{Vzeta}), 
i.e., $\zeta_1>0$, always becomes 
singular during collapse and so  we need two real positive roots for the
above equation (\ref{Vzeta}) for a possibility of rebounce.  The
region between the two real positive roots is forbidden region.
So if we start on the right side, the collapsing shells bounce back and
then we have continuous expansion.
Since one of the real root has to be negative and the region between two
real positive roots is forbidden, it is not possible to have
oscillating solutions in these spacetimes.

In the case of $\Lambda =0$ it is well-known that we cannot have 
a rebounce and the collapse necessarily results in a singularity. 
The cubic then reduces to a linear equation and the solution is given as,
$R=R_{max}(r)=-F/f$. So only in the case $f<0$ we can have 
$\dot R =0$ for positive $R$. This corresponds to the maximum possible
physical radius for a given shell i.e. even if we start with initial
expansion a given shell will reach the maximum radius $R_{max}(r)$ and then
it will recollapse.


In the case when $\Lambda<0$,
one and only one root is always positive.  The other two roots are
negative if $9F^2< -({4f^3}/{\Lambda)}$, or else
they are complex conjugates. Therefore any initial configuration becomes
singular in this case. The real positive root in this case gives an 
upper bound on the radius $R=R_{max}(r)$ of a shell labeled $r$. 
This upper bound occurs as
the negative (attractive) contribution from 
the $\Lambda$ term keeps on
increasing with an increasing $R$, while the contribution from gravitational
attraction keeps on decreasing and so at some point for any value $f$
the attraction due to $\Lambda$ starts dominating. Hence, even if 
we have an initially expanding configuration, finally we always 
must have collapse in this case.

For $\Lambda > 0 $ and $f(r)\ge 0$ one can easily see 
that we can never have $\dot{ \zeta}^2=\dot R^2/r^2 =0$, 
so singularity always forms 
if we start from an initial collapse. 

For the  case $\Lambda > 0 $ and $f(r)<0$, one root of the 
cubic above is always negative. If 
\begin{equation}
F^2> -{4f^3\over 9\Lambda},
\label{nobounce1}
\end{equation}
then  the other two
roots are complex conjugates. So the singularity always forms in such a 
case in initially collapsing configuration. On the other hand, if
the initial data is such that  
$$
F^2< -{4f^3\over 9\Lambda},
$$
then the other two roots are  real positive. Let us denote them by 
$\zeta_1$ and $\zeta_2$ with 
$\zeta_1<\zeta_2$. The region between the two roots is forbidden.
The entire space of allowed
dynamics is given by the two disjoint regions
$[0, \zeta_1]$ and $[\zeta_2,\infty]$. If the initial scale 
factor $\zeta_0$ lies in the first section, i.e. if $\zeta_0<\zeta_1$,
then we always have
the singularity as the end point of collapse. 
Here $r\zeta_1$ represents the upper bound for the physical
radius of a shell in this region.  If $\zeta_0$ lies in the second section,
i.e. $\zeta_0>\zeta_2$, then we will have a rebounce from the
initial collapsing configuration. After the rebounce the physical radius
of the shell keeps on increasing forever. There is no upper limit for the
maximum value of $\zeta$ in this region and $r\zeta_2$ gives the lower bound
for the physical radius of a shell, i.e. the shell rebounces at 
$R=r\zeta_2$.

>From the above discussion we can see that we can have rebounce only
in the case when $\Lambda>0$ and $f<0$, and when the following two 
conditions are satisfied,
\begin{equation}
F^2< -{4f^3\over 9\Lambda},
\label{bounce1a}
\end{equation}
and if $\zeta_0>\zeta_2$. With our scaling, this later condition
can be written as,
\begin{equation}
1> 2{\bigg (} -{f\over r^2\Lambda}{\bigg
)}
\cos{{\bigg[}{1\over 3}\arccos{\sqrt{-{9F^2\Lambda\over 4f^3}}}
{\bigg ]}}.
\label{bounce2}
\end{equation}
The tables \ref{TLLf1} and \ref{TLLf2} give a systematic classification
of the various possibilities we discussed above.

The physical quantities like central density, $\rho_c=F_0/\xi^3(t)$, and
curvature scalars
\[
R^{ijkl}R_{ijkl}=12\left(\frac{{\dot \zeta}^4 + {\zeta}^2{\ddot
\zeta}^2}{{\zeta}^4} \right),
R^{ij}R_{ij}=12\left(\frac{{\dot \zeta}^4 + \zeta {\dot \zeta}^2 
{\ddot \zeta} + {\zeta}^2{\ddot \zeta}^2}{{\zeta}^4}\right),
{\cal R} = 6\left(\frac{{\dot \zeta}^2 +  \xi {\ddot \zeta}}{{\zeta}^2}
\right)
\]
stay finite as $\zeta(t,r)>0$ for the regular models. 

\begin{table}[t]
\begin{tabular} {|c|c|c|c|}
\hline

$\Lambda $ & $f>0$ & $f=0$ & $f<0$ \\
\hline

$\Lambda <0$ & Closed solution,  &  Closed solution, &
 Closed solution, \\

 & No rebounce & No rebounce & No rebounce \\
\hline

$\Lambda =0 $ & Open (hyperbolic) solution, & Marginal 
(parabolic) solution, & Closed (elliptic) solution, \\
 & No rebounce & No rebounce & No rebounce \\
\hline

$\Lambda >0 $ & Open solution, & Open solution, &
Various cases occur \\
 & No rebounce & No rebounce &  as in table \ref{TLLf2}. \\
\hline

\end{tabular}

\caption{The various types of dust solutions with $\Lambda $}

\label{TLLf1}
\end{table}
\begin{table}
{\begin{center}
\begin{tabular}{|c|c|c|}
\hline

$ F^2> -\frac{4f^3}{9\Lambda}$& $-$ & Open solution, \\
& &  No rebounce\\
\hline

 $F^2< -\frac{4f^3}{9\Lambda}$ & $\zeta_0<\zeta_1 \; \Rightarrow $ 
& Closed solution, \\
&$ 1< 2{\bigg (} -{f\over r^2\Lambda}{\bigg
)}
\cos{{\bigg[}{4\pi\over 3}+ {1\over 3}\arccos{\sqrt{-{9F^2\Lambda\over 4f^3}}}
{\bigg ]}}$
 & No rebounce \\
\hline

$F^2< -\frac{4f^3}{9\Lambda}$ & $\zeta_0>\zeta_2  \; \Rightarrow $ & 
Open Solution,\\
& $1> 2{\bigg (} -{f\over r^2\Lambda}{\bigg
)}
\cos{{\bigg[}{1\over 3}\arccos{\sqrt{-{9F^2\Lambda\over 4f^3}}}
{\bigg ]}}$
  & Rebounce\\
\hline

\end{tabular}

\end{center}
}
\caption{The various types of dust solutions with $\Lambda>0$ and $f<0$} 
\label{TLLf2}
\end{table}

There are several features here which are worth noting and have 
interesting physical significance as far as the dynamics of collapse
is concerned, and which illustrate the effects a non-vanishing 
cosmological term may have towards determining the final fate of a 
collapsing cloud of matter. Firstly, with a negative value of the
cosmological term all the solutions become closed and singularity
always forms in a future even if we start with initial expansion. 
This is to be expected because such a value will only contribute
in a positive manner to the overall gravitational attraction of matter
and just acts as a constant positive energy field helping the collapse.
Next, as illustrated by the first line of Table \ref{TLLf2}, 
there is a range of 
initial data where the collapse necessarily ends in a singularity despite 
however large positive value of the cosmological term. This is contrary
to the belief sometimes expressed that a positive cosmological constant
can always cause a bounce provided it is sufficiently large in magnitude.
Finally, it is clear from above that there can be a rebounce only if
the initial density is sufficiently low for a given positive value
of $\Lambda$. This is so because the cosmological term becomes
dominant with increasing distances, and gravity dominates at higher
densities. Thus as the cloud is more disperse (lesser densities but bigger 
size), more is the effect of the cosmological term.

\subsection{The $f=0$ case}

As shown in the previous consideration, while a bounce and regular 
solution occurs for a specific range of the initial data,
for a majority of the regular initial data space the collapse results
into a spacetime singularity where densities and curvatures blow
up. While for $\Lambda=0$ case we know in detail the structure of this 
singularity, and when it will be naked or covered. 
We would like to understand here the effects of a non-zero $\Lambda$ 
towards the structure of the singularity forming in such a collapse.

To study these features for the dust collapse models,
we analyze now explicitly the case $f(r) = 0$.
While we have chosen $f=0$ for simplicity and clarity of the
considerations, one expects quite similar behaviour in other cases also.
The equation~(\ref{dyna}) can now be written as,
\begin{equation}
t-t_{c}(r) = \pm {\int ({\frac{F(r)}{R} + 
\frac{\Lambda}{3}R^2})^{-1/2}}dR,
\label{elliptic}
\end{equation}
where $t_{c}(r)$ is an integration function which 
it represents  the time at which a given shell, $r$, becomes 
singular, i.e.  $R(t_{sf}(r),r)=0$.
The positive or negative sign respectively corresponds to the
expanding and collapsing branches of the solution. 
In what follows, we 
consider only the negative sign because we are considering clouds which are
collapsing, with $\dot R <0$ initially. The integral~(\ref{elliptic}) can be
written as an infinite series in $R$ near the centre as below, 
\begin{equation}
t - t_{c}(r) =-\frac{2}{3}\frac{R^{3/2}}{\sqrt{F(r)}} \left[ 1 +
	 \sum_{m=1}^{\infty} {\frac{(-1)^m (2m-1)!!}{ 2^m (2m+1) m!}
\left(\frac{\Lambda R^3}{3F(r)} \right)^m} \right] .
\label{solution}
\end{equation} 
Using the scaling freedom in our solution, we set $R(t=0,r)=r$ 
which determines  $t_{c}(r)$ as
\begin{equation}
t_{c}(r) = t_{sf}(r) = \frac{2}{3} \frac{r^{3/2}}{\sqrt{F(r)}} \left[ 1 +
\sum_{m=1}^{\infty} \frac{(-1)^m (2m-1)!!}{2^m (2m+1) m!}
\left( \frac{\Lambda r^3}{3F(r)} \right)^m \right] .
\label{sincurve}
\end{equation}

>From the above expressions we get,
\begin{equation}
R'= \frac{F'}{3F}R + \left( 1- \frac{rF'}{3F} \right) \left(
\frac{3F+ \Lambda R^{3/2} }{ 3F+ \Lambda r^{3/2}}
\right)^{1/2} \left( \frac{r}{ R} \right)^{1/2}.
\label{Rprime}
\end{equation}  
Therefore, the condition $R'> 0$, implying no shell-crossings, can be
satisfied if $F'> 0$ and $1-{rF'/3F} > 0$. This means that the total mass 
inside a shell and the matter density are increasing and decreasing 
functions of $r$ respectively, as we move away from the centre. 
The weak energy condition, implying positivity of energy density, 
guarantees that mass is an increasing function of $r$. Also, for
any realistic density distributions, it would be physically reasonable
that the density is higher at the center, decreasing away from the
center. Thus we shall work with decreasing
density profiles. Therefore, no shell-crossings occur in the spacetime
for our choice of initial data, before the occurrence of the
central shell-focusing singularity.

As we will show in the next Section, the behaviour of collapsing shells
near the centre would depend only on the first non-vanishing derivatives of
the density and velocity profiles near the same. Therefore, the local
visibility conditions are unaffected by the boundary conditions such as
the initial choices of the mass function and the actual value of$\Lambda$. 
On the other hand, the global behaviour of the trajectories coming out 
from the singularity can change due to the 
addition of a cosmological term.

\section{The structure of the singularity}

In this Section we will analyze the nature of
the central singularity. In what follows we use the scheme, as 
developed earlier by Joshi and Dwivedi~\cite{JoshiPRD}, which
gives us a necessary and sufficient condition for the local visibility
of the singularity. The main idea here is to see if we can have 
ORANGEs in the spacetime, meeting the singularity in their past with a 
well-defined real positive tangent vector in a suitable plane.

The equation for the ORANGEs in spacetime (\ref{metric}), 
for $f(r) = 0$ is
\begin{equation}
\frac{dt}{dr} = R' .
\end{equation}
For convenience this can be written in the $(u,R)$ plane as,
\begin{equation}
\frac{dR}{du} = \frac{R'}{ \alpha r^{\alpha -1}} \left(1 -\sqrt{\frac{F}{R} +
\frac{\Lambda}{3}R^2} \right) ,
\label{null}
\end{equation}
where $u = r^{\alpha}$, and $\alpha \geq 1$ 
is a constant to be determined later. 
>From equation (\ref{null})
it is clear that there are no ORANGEs from the non-central part
of the singularity curve
because the first term under square-root goes to $-\infty$ as $t$ 
approaches $t_{sf}(r)$. This means that we cannot have ORANGEs as 
solutions to above
equation near the non-central singularity, and hence it is only the central 
singularity which can be possibly visible.

We define $X=R/u$, to check if we can have a well-defined tangent for the 
equation~(\ref{null}), at $r=0,\; R=0$ in the limit of $t$ approaching 
the singular epoch $t_{s}(0)$.  Using the l'Hospitals rule we get,
\begin{equation}
X_0= \lim_{u\to0,R\to0} {R\over u} = \lim_{u\to0,R\to0} {dR\over du} =
 \lim_{u\to0,R\to0} {R'\over \alpha r^{\alpha -1}} {\bigg (} 1 -\sqrt{
{F\over R} +
{\Lambda \over 3}R^2} {\bigg )} = U(X_0,0)
\label{root1}
\end{equation}
where the subscript $0$ denotes the value of the quantities at $u=0$. 
The constant $\alpha$ can be uniquely fixed by demanding that 
${R'/ r^{\alpha -1}}$ is non-zero finite~\cite{JoshiPRD}. 
In the case that we are discussing,  ${R'/ r^{\alpha -1}}$ remains finite 
if we choose $\alpha=1+2n/3$~\cite{JoshiPRD}, and the above equation
~(\ref{root1}) can be written as,
\begin{equation}
{1\over \alpha} {\bigg [} X- {nF_n\over 3F_0} {1\over \sqrt X} {\bigg ]}
{\bigg [} 1 + \sqrt{\lambda_0 \over X} {\bigg ]}=0,
\label{roots}
\end{equation}
where  $\lambda = F/u$. If this equation has a real positive root $X_0$ 
then one will have at least one null geodesic coming out of the 
singularity at 
$R=0,u=0$ with the root $X=X_0$ as a tangent in $(u,R)$ plane.

When $n < 3$, we have $\alpha < 3$  and therefore $\lambda_0=0$,
and the above equation reduces to,
\begin{equation}
X_0^{3/2}=-\frac{F_n}{ 2F_0 \sqrt{1+\Lambda F_0/3}}
\label{secroot}
\end{equation}
which always has a real positive root. Apart from an additional $\Lambda$
term, this equation is analogous to the corresponding equation obtained
for the LTB models, and
reduces to the same for $\Lambda=0$. Therefore, when either the first or
the second derivative of density is non-zero the singularity is always, at
least locally, visible. The addition of cosmological term changes only 
the value of the tangent ($X_0$) to the ORANGES, but not the visibility
property itself. Thus the corresponding
dust naked singularity spectrum is stable to the addition of a positive
cosmological constant.

In this case, like the earlier studies on LTB models, 
the smaller root will be along the apparent
horizon direction and a family of geodesics will come out along this 
direction~{\cite{GRG}}. In this case we put $\alpha=3$ and the first term 
in the equation (\ref{null}) blows up and second term goes to zero such that
the product is $F_0$.

For the value $n=3$, which corresponds to the critical case in LTB 
models~\cite{Singh}, we get $\lambda_0= F_0$. Introducing 
$X= F_0 x^2$ and $\xi_{\Lambda} = {F_3}/ F_0^{5/2} \sqrt{1+\Lambda F_0/3}$
the root equation becomes,
\begin{equation}
2x^4+x^3-\xi_{\Lambda} x +\xi_{\Lambda} =0 .
\end{equation}
This equation is similar to the corresponding case in the LTB models,
with the modification in the definition of $\xi$. From the theory of
quartic equations, this equation admits a real positive root for
$\xi_{\Lambda} < \xi_{crit} = -(26+15{/\sqrt{3}})/2 $. Therefore, for a
given central density $F_0$, and the inhomogeneity parameter as
given by $F_3$, the
naked dust singularity can be partly covered by a positive $\Lambda$,
as we get an
additional positive term  in the denominator in $\xi_{\Lambda}$. 
But it is interesting to note that, however large, a finite $\Lambda$
term cannot completely cover the corresponding visible part of dust models.

In a similar manner we can see that a negative $\Lambda$ will open-up
some covered part in the dust spectrum.
(As such our calculation does not use the positivity of $\Lambda$ anywhere,
so they go through even for negative $\Lambda$.)

For all values $n>3$ we have $\lambda_0 = \infty$, and we cannot 
have a real positive root in these cases and the final singularity 
is safely hidden behind the event horizon.

\section{Conclusion}

Studying gravitational collapse of a dust cloud with a non-zero
value of the cosmological term we observed that black holes, naked 
singularities and even globally regular solutions can develop as
the final outcome of collapse. 
Each of these outcomes is determined by 
the choice of initial parameters, given in terms of the density and
velocity profiles of the cloud. 
Though for simplicity we restricted to $f(r) = 0$ case 
while analyzing the structure of the singularity,
the results can be extended to the general case. 
We have also shown here that
it is possible to cover a part of the naked singularity spectrum, in the
corresponding critical branch of solutions in the LTB models, with
the introduction of a positive cosmological constant. An interesting 
conclusion that emerges is that the existence of naked singularity remains 
stable to the introduction of a cosmological term in Einstein
equations. These results are of interest in that they allow us to
understand the implications of a non-zero $\Lambda$ towards
the final outcome of gravitational collapse, in view of the recent 
observational claims about a non-vanishing cosmological constant. 

We acknowledge the support from
University of Basque Country Grants UPV122.310-EB150/98, UPV172.310-G02/99
and the Spanish Ministry Grant PB96-0250.

\end{document}